\documentclass[prl,twocolumn,nobibnotes,floatfix,showkeys,superscriptaddress]{revtex4}

\usepackage{latexsym,amsmath,amssymb,amsfonts,mathbbol,graphicx,color}
\usepackage{dcolumn}% Align table columns on decimal point
\usepackage{bm}% bold math

\begin{document}

\title{Storing Small Photonic Cluster States in a Dephasing Environment}

\author{Yaakov S. Weinstein}
\author{Gerald Gilbert}
\affiliation{Quantum Information Science Group, {\sc Mitre},
260 Industrial Way West, Eatontown, NJ 07224, USA}

%\date{\today}
\begin{abstract}
We consider the effects of decoherence on the entanglement of photonic cluster states.
Large photonic cluster states can be built by fusing together smaller photonic cluster states via 
probabilistic fusion operations. For this construction process it is necessary to 
store these smaller cluster states in some way so as to have them available for attempted 
fusion operations. While in storage the photonic cluster states may undergo dephasing. 
The affects of dephasing on small, primitive cluster states is explored here with the aim of 
determining how to locally rotate the qubits of the cluster state so as to lose the 
least amount of entanglement due to the dephasing process. 
\bigskip
\keywords{cluster states; entanglement; decoherence}
\end{abstract}

\maketitle

%\doi{10.1080/0950034YYxxxxxxxx}
% \issn{1362-3044}
%\issnp{0950-0340} \jvol{00} \jnum{00} \jyear{2008} \jmonth{10 January}

\section{Introduction}

Cluster states are highly entangled states that can be used as
a resource for universal quantum computation via single qubit measurements only 
\cite{BR1,BR2,BR3}. To create a cluster state each of the initially unentangled qubits must be rotated 
into the state $|+\rangle = \frac{1}{\sqrt{2}}(|0\rangle + |1\rangle)$. Desired pairs of qubits are then 
entangled by applying control phase (CZ) gates between them. In a graphical picture of a 
cluster state, qubits are represented by circles and pairs of qubits that have
been entangled via a CZ gate are connected by a line.  A cluster state
with qubits arranged in a two-dimensional lattice, such that each qubit
has been entangled with four nearest neighbors, suffices for universal QC.

After construction, any quantum computational algorithm can be implemented 
via the cluster state using only single-qubit measurements along axes in the $x$-$y$
plane. These processing measurements are performed by column, from left 
to right, until only the last column is left unmeasured. The last column 
contains the output state of the quantum algorithm which can be extracted 
by a final readout measurement. 

A possible experimental venue for cluster states is photonics. 
Nielsen noted \cite{N} that a photonic cluster state quantum computation may be more efficient 
than a circuit model quantum computation if certain techniques from linear optics 
quantum computation, were used to build the photonic cluster. Browne and Rudolph \cite{BR} 
refined this idea replacing Nielsen's construction method with simpler `fusion' operations. 
A number of additional methods for constructing clusters have been suggested \cite{DR,CCWD} 
and small photonic cluster states have been experimentally implemented \cite{Zeil,K,Pan}. An 
improvement of the Browne and Rudolph scheme, that utililzes cluster state equivalences \cite{HEB} 
to remove the need to perform costly Type II fusion operations, was discussed in \cite{us}.

As with any attempted implementation of quantum computation attention must be paid to the effects of 
decoherence on the system of interest. In photonic cluster state quantum computation decoherence can 
disturb the implementation during two separate experimental phases: the construction of the cluster state, 
and the measurement of the cluster state qubits, {\emph i.e.} the implementation of the desired algorithm. 
In the construction phase we assume a constant source of two qubit photonic cluster states, such as 
a spontaneous paramtric downconversion source. The two qubit clusters are `fused' via Type I operations
to create higher qubit cluster states. However, these cluster building blocks as well as the resulting 
cluster may need to be stored so as to be available for subsequent fusion operations or algorithm implementation. 
The storage of these two, three or more qubit clusters is necessarily in a decohering environment 
(here the decoherence is assumed to be dephasing due to birefringence). We would like to utilize 
the freedom in performing one qubit rotations, a relatively easy task in photonic quantum computation, to arrange 
that the cluster state be stored in a way most robust against decoherence. We measure this immunity 
based on the amount of entanglement lost due to the decoherence and explore cluster states of two, 
three, and four qubits. 

We define a representation as the exact wavefunction of an $n$-qubit cluster state. Different 
representations of the same $n$-qubit cluster state differ by single qubit rotations. 
What we call the basic representation is the output of the construction method 
mentioned above, placing the qubits in the $|+\rangle$ state and applying desired controlled phase 
operations. These states have density matrices whose elements all have the same absolute value. 
Hadamard gates are then used on individual or multiple qubits to change between different 
state representations.

We assume there are no interactions between the qubits and thus the only dynamics in the system is 
the dephasing of the storage unit. This dephasing is fully described by the Kraus operators
\begin{equation}
K_1 = \left(
\begin{array}{cc}
1 & 0 \\
0 & \sqrt{1-p} \\
\end{array}
\right); \;\;\;\;
K_2 = \left(
\begin{array}{cc}
0 & 0 \\
0 & \sqrt{p} \\
\end{array}
\right)
\end{equation} 
where we have defined the dephasing strength $p$. When all $n$ 
qubits undergo dephasing we have $2^n$ Kraus operators each of the form 
$A_l = (K_i\otimes ... \otimes K_{\ell})$ where 
$l = 1,2,...,2^n$ and $i,...,\ell = 1,2$. 
All of the below calculations
are done with respect to $p$, where the exact behavior of $p$ as a function 
of time is left ambiguous so as to accomodate various possible dephasing behaviors. 
As an example, we may assume $p = 1-e^{-\kappa\tau}$ where $\tau$ is time and 
$\kappa$ is the decay constant. In this case, off diagonal terms of the density matrix
decay as a power of $e^{-\kappa t}$ and thus go to zero (i.~e.~$p\rightarrow 1$) only at 
infinite times. We also assume equal dephasing for all qubits. 

For the sake of consistency, we will always use negativity based entanglement measures, $N$, 
regardless of the number of qubits in the system of interest. We define the negativity simply as the most negative 
eigenvalue of the parital transpose of the density matrix \cite{neg}. When the system consists 
of more than two qubits, there are a number of inequivalent forms of the negativity: the partial 
transpose may be taken with respect to any single or any permutation of multiple qubits. We 
use a subscript to denote with which qubits the partial transpose has been taken. We also make
use a the tri-partite negativity, $N^3$, a tri-partite entanglement 
measure for mixed states which is simply the third root of the product of the 
negativities with respect to each of three qubits \cite{SGA}.

\section{Two Qubit Clusters} 
\begin{figure}[t]
\begin{center}
\includegraphics[width=6.5cm]{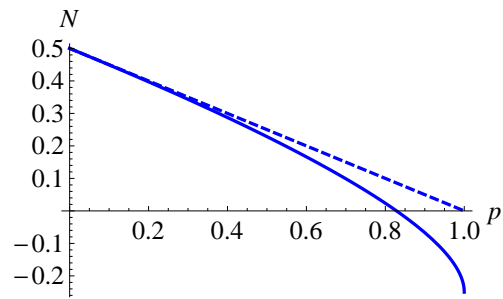}
\caption{\label{TwoQubit} (Color online)
Entanglement, $N_1$, as a function of dephasing strength for two locally equivalent representations
of the two-qubit cluster state, the basic representation (solid line) and the GHZ state (dashed line).
Note that only the first representation undergoes a complete entanglement sudden death.
}
\end{center}
\end{figure}

For two qubits we look at two locally equivalent representations of the cluster state. The basic 
representation is the state $\frac{1}{2}\left(|00\rangle+|01\rangle+|10\rangle-|11\rangle\right)$,
and the representation for which a Hadarmard is performed on one of the qubits: 
$\frac{1}{\sqrt{2}}\left(|00\rangle+|11\rangle\right)$, a two qubit Greenberger-Horne-Zeilinger (GHZ)
state. The decay of entanglement due to dephasing
for these two representations are shown in Fig. \ref{TwoQubit}. For strong dephasing the the first 
representation undergoes a complete loss of entanglement, entanglement sudden death (ESD) \cite{ESD}, at
$p = 2(\sqrt{2}-1) \simeq .828$. While some amount of entanglement is always found in the second , GHZ,
representation. In addition, the amount of entanglement present in the GHZ state is always higher than
that of the basic representation. This implies that for purposes of storage in a dephasing environment 
the GHZ state should be the state representation of choice. 

\section{Three Qubits}
For three qubit cluster states both the number of state representations and negativity entanglement measures
increase. We explore the entanglement decay for three different representations. For each 
representation we look at three negativity measures $N_i$, the partial transpose taken with respect to each of 
the qubits, $i = 1,2,3$. In the basic representation all elements of the density matrix have equal absolute value. 
Each of the negativity values for this representation undergo ESD. For negativities where the partial transpose is taken with 
respect to the first or third qubit ESD is exhibited at dephasing strength $p = 2(\sqrt{2}-1) \simeq .828$.
For the negativity taken with respect to the second qubit this occurs at $p \simeq .913$. Applying single qubit 
Hadamard rotations on the first and third qubit transforms the basic representation into a three qubit GHZ state. All 
three negativity measures for the GHZ state behave the same way, and lose their entanglement only in the limit 
as $p \rightarrow 1$. If instead, a Hadamard rotation is applied only the third qubit of the basic representation 
(or to all three qubits) the density matrix has 16 non-zero elements of equal magnitude. When the partial transpose of this 
third representation is taken with respect to the first or second qubit, the negativity never exhibits ESD and approaches zero
more slowly than the negativities associated with the GHZ state. If, however, the partial transpose is taken with 
respect to the third qubit, the entanglement disappears at a weaker dephasing strength than any we have yet seen 
$p \simeq .704$.

\begin{figure}
\begin{center}
\includegraphics[width=6.5cm]{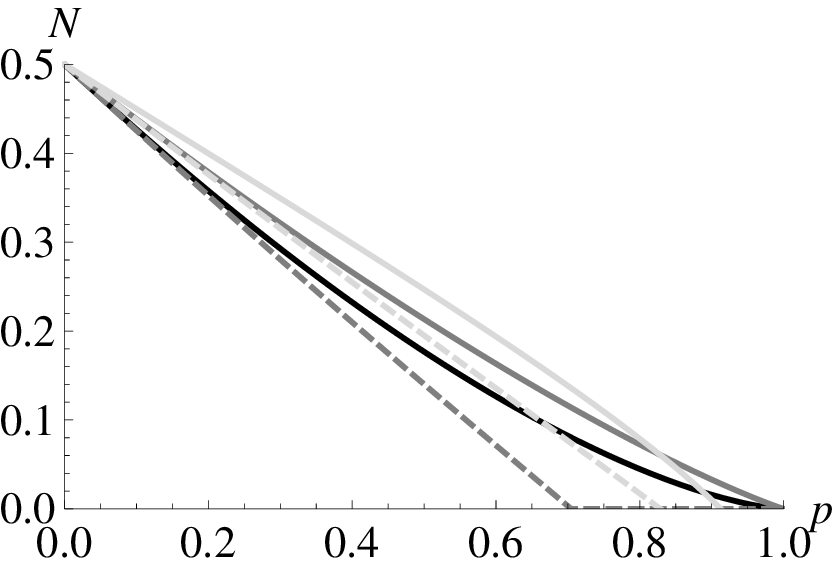}
\includegraphics[width=6.5cm]{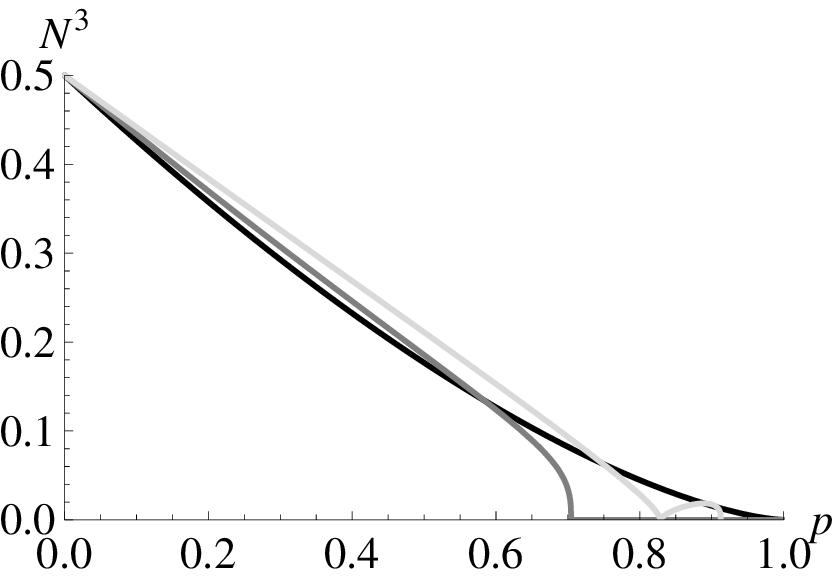}
\caption{\label{ThreeQubit} (Color online)
Possible negativity and tri-negativity decays for different representations of the three qubit cluster state 
as a function of dephasing strength, $p$. Left: Five possible decays of negativity for the GHZ state (black),
the third representation (gray) and the basic representation (light gray). For the third representation there 
are two different negativity behaviors, $N_1$ and $N_2$ (solid line), and $N_3$ (dashed line). Similarly, the 
basic representation exhibits two negativity decay behaviors $N_2$ (solid line), and $N_1$ and $N_3$ (dashed 
line). Right: Decay of tri-negativity for each of the three cluster state representations. 
}
\end{center}
\end{figure}

Fig. \ref{ThreeQubit} (left) plots the five independent behaviors of the negativity for the three qubit cluster
state and leaves us with interesting criteria for choosing the appropriate representation for robust storage. 
At low values of $p$ the most robust entanglement is $N_2$ for the basic representation of the cluster state. The other  
negativities, $N_1$ and $N_3$ are slightly less robust but are still higher then almost all other entanglements (especially at 
very low values of $p$). At high values of $p$, no GHZ negativities disappear, though the entanglement of the GHZ 
representation is generally the second lowest of all the three qubit negativity measures. The third representation
has a negativity that is always the lowest and two negativites that are second highest or highest (at high $p$
values). The figure (right) also shows the tri-partite negativity, $N^3$ for each of the three representations. This 
purely tri-partite negativity decays the slowest for the basic representation of the three qubit state until
high values of $p$. It, and the tri-partite negativity for the third representation exhibit 
ESD at the same values for which ESD of the negativity is exhibited. $N^3$ for the GHZ representation however, 
does not exhibit ESD. 

\section{Four Qubits}
A four qubit cluster allows the implementation of an arbitrary single qubit rotation in the cluster state
quantum computing paradigm. The decay of entanglement as a function of decoherence strength for two 
representations of the four qubit cluster state was explored in \cite{YSW}. In that context the entanglement
disappearance was compared to the fidelity of the attempted rotation. Here we explore five represenations 
of the four qubit cluster to determine which is best to use when storing the state in a dephasing environment.
We note that, unlike two and three qubits, the four qubit cluster state cannot be rotated into a GHZ state. 
A four qubit state also allows for many more negativity metrics as now inequivalent partial transposes can be 
taken with respect to both one and two qubits. 

\begin{figure}
\begin{center}
\includegraphics[width=6.5cm]{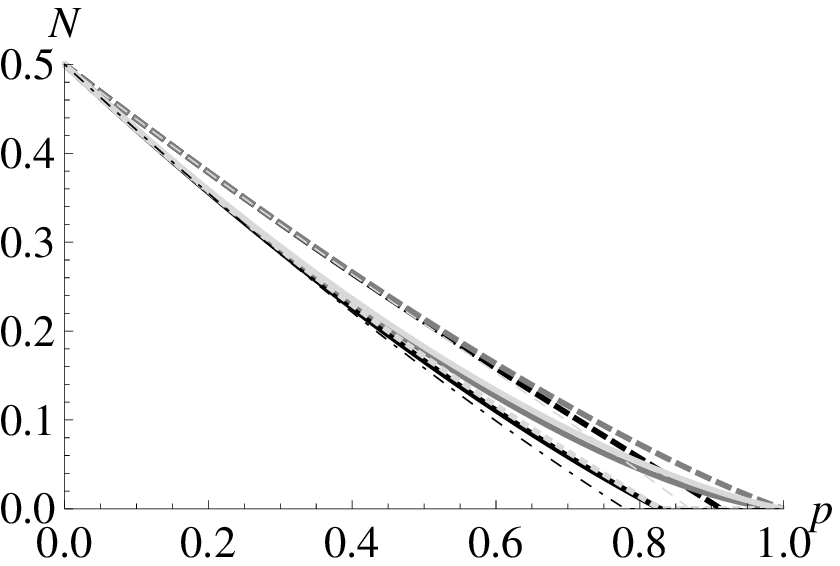}
\includegraphics[width=6.5cm]{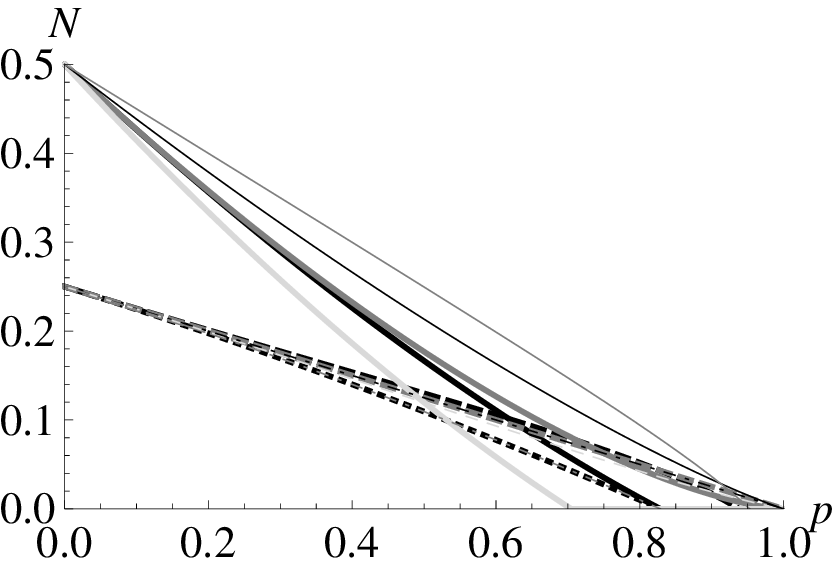}
\caption{\label{FourQubit} (Color online)
Negativity for different representations of the four qubit cluster state with partial transpose taken
with respect to one (left) and two (right) qubits. In order of representations the entanglement is shown
by lines that are black, gray, light gray, thin black and thin gray. The different patterns (solid, dashed,
dotted and chained) represent different negativities. Left: Negativities with partial transpose taken with 
respect to single qubits, $N_i$. The negativities most robust against entanglement are $N_3$ for the second 
representation (gray dashed) which exhibits the same behavior as $N_3$ and $N_4$ of the fourth representation.
This is followed by $N_2$ for the basic representation which is equivalent to all four of the single qubit 
negativities of the fifth representation.  
The least robust is the $N_2$ for the fourth representation. A number of the single qubit negativities 
exhibit ESD at value $.75 < p < .92$. Right: Negativites with partial transpose taken with respect to two
qubits, $N_{ij}$. The most robust of these negativities is $N_{34}$ of the fifth representation followed
by $N_{34}$ of the fourth representation. The former of these exhibits ESD while the latter does not. The 
least robust of these negativities (solid light gray)is $N_{34}$ of the third representation. Once again
many of the negativity measures exhibit ESD for a range of dephasing values $.7 < p < .95$.  
}
\end{center}
\end{figure}

We find five representations that exhibit different entanglement decay behavior. These are: the basic representation,
Hadamards applied to qubits one and three, a Hadamard applied only to qubit one, Hadamards applied to qubits one, two,
and three, and Hadamards applied to qubits two and three. Different behaviors for the negativity with partial 
transpose taken with respect to one $N_i$ and two $N_{ij}$ qubits are shown for the five representations in 
Fig. \ref{FourQubit}. 

Selecting the proper storage representation given the jungle of data is difficult as there is no clear 
representation that is always most robust against dephasing. However, overall the fifth representation seems to be
always close to the most robust. A useful approach may be to determine in advance what the cluster will be used
for and perform appropriate simulations to calculate which decohered state representation will be best for that
particular purpose. This approach is similar to the simulations carried out in \cite{YSW} in which fidelities 
were calculated for arbitrary one qubit cluster state rotations using as a resource two different representations 
of a four qubit cluster state.

\section{Conclusion}

In this paper we have explored the effects of dephasing on two, three, and four qubit cluster states. The point 
of this exercise is to determine which state representations entanglement is most robust when stored in a decohering 
environment. Such storage
is necessary in attempts to build photonic cluster states. We have found that for two qubits the GHZ representation
is more robust against decoherence than the basic representation. For cluster states with three and four qubits 
the proper choice of representation will depend on how strong the decoherence is and what entanglement is most
important to preserve. This will most likely depend on the particular role of this cluster state. 

It is a pleasure to acknowledge support from the MITRE Innovation Program under MIP grant \#20MSR053.

\end{document}